# Quantum Computing Assisted Deep Learning for Fault Detection and Diagnosis in Industrial Process Systems

Akshay Ajagekar, Fengqi You[*]

Cornell University, Ithaca, New York 14853, USA

## Abstract

Quantum computing (QC) and deep learning techniques have attracted widespread attention in the recent years. This paper proposes QC-based deep learning methods for fault diagnosis that exploit their unique capabilities to overcome the computational challenges faced by conventional data-driven approaches performed on classical computers. Deep belief networks are integrated into the proposed fault diagnosis model and are used to extract features at different levels for normal and faulty process operations. The QC-based fault diagnosis model uses a quantum computing assisted generative training process followed by discriminative training to address the shortcomings of classical algorithms. To demonstrate its applicability and efficiency, the proposed fault diagnosis method is applied to process monitoring of continuous stirred tank reactor (CSTR) and Tennessee Eastman (TE) process. The proposed QC-based deep learning approach enjoys superior fault detection and diagnosis performance with obtained average fault detection rates of 79.2% and 99.39% for CSTR and TE process, respectively.

## 1. Introduction

Fault detection and diagnosis has been an active area of research in process systems engineering due to the growing demand for ensuring safe operations and preventing malfunctioning of industrial processes by detecting abnormal events [1, 2]. Furthermore, the advent of chemical plant accidents causing tremendous environmental and economic losses provide an extra incentive to develop process monitoring techniques that effectively assure process safety and product quality in complex chemical process systems. Data-driven approaches often termed as multivariate statistical process monitoring methods have attracted significant attention and have been widely applied to monitor industrial processes [3-5]. Such methods rely on historical process data and rarely require detailed knowledge of the governing physical models of the continuous or batch processes [6], thus making them relatively easier to implement in process control and operations [7].

---

[*]Corresponding author. Phone: (607) 255-1162; Fax: (607) 255-9166; E-mail: fengqi.you@cornell.edu

Quantum computing (QC) based applications have been gaining traction recently due to their unique capabilities with a significant portion of its presence perceived in the area of optimization with applications in operational planning [8], molecular design [9, 10], process scheduling and operations [10, 11], logistics optimization [10, 12], and energy systems [13]. The randomness and uncertainty inherently associated with QC operations, subject to internal magnetic fields, thermal fluctuations, and other noise sources, could be a hindrance to optimization applications. However, this non-ideal behavior can be exploited to develop efficient statistical machine learning techniques. QC-enhanced machine learning techniques have been proposed for data fitting [14], pattern recognition [15], generative machine learning [16], handwriting recognition [17], and quantum recommendation systems [18]. These QC-based data-driven techniques can also be used in process control and monitoring for industrial processes. Quantum advantages offered by QC in terms of speed and method of operation could benefit fault monitoring in complex process systems where swift and precise fault detection is desired. However, the applicability of QC-based techniques is limited due to the commercially available quantum computers facing several limitations like low number of quantum bits (also termed as qubits), limited connectivity, and lack of quantum memory. As a result, integrating QC-enhanced learning techniques with classical machine learning algorithms to overcome such limitations becomes necessary and is a promising approach for process monitoring.

The applicability and capacity of some basic classical data-driven methods in industrial process monitoring such as principal component analysis (PCA), partial least squares (PLS), independent component analysis (ICA), and fisher discriminant analysis (FDA) has been extensively studied [19, 20]. PCA and FDA are dimensionality reduction techniques that can be used to detect faults and discriminate among classes of data by describing the trends in historical data through lower dimensional representations [21, 22]. PLS and ICA are other powerful multivariate statistical tools widely used for fault detection and diagnosis [23, 24]. Monitoring techniques based on these methods face some limitations which directly affect their anomaly detection efficiency in complex process systems. PCA-based methods do not take into account the temporal correlations between process data and information between classes when determining the lower dimensional representations. FDA requires control limits for fault detection devised from the assumption that the measurement signals follow a multivariate Gaussian distribution which may raise false alarms. It is often difficult to interpret the independent latent variables in PLS with a possible risk of overfitting. Several new variations of the basic data-driven monitoring methods have also been proposed and applied to fault detection and diagnosis in industrial processes [25-28]. However, a large portion of these analytical approaches are limited to linear and some specific nonlinear models. Also, the inherent nonlinear nature of complex process systems renders the use of such methods inefficient due to misclassification of large portion of the process data. Nonlinear classification techniques like support vector machine (SVM) improve the fault classification performance for highly overlapped data. However, the corresponding model complexity increases with the process data dimensions [22]. The extent of complex nonlinearities and correlations present between the process data make it difficult for these classical data-driven methods to generalize to all complex process systems, restraining their applicability in practical situations.

The ability of artificial neural networks to approximate nonlinear relationships between the process data and process states by generalizing the knowledge can be successfully applied to diagnose faults in complex chemical process systems [29-32]. However, in some instances their generalization to multiple faults is not always successful. Recently, deep learning has become a promising tool for smart fault diagnosis due to powerful techniques like auto-encoder (AE) [33], restricted Boltzmann machine (RBM) [34] and convolutional neural network (CNN) [35, 36]. Such deep learning models extract multiple levels of abstraction from normal and faulty data, allowing them to achieve high classification accuracy. The increasing complexity of industrial process systems requires deeper and more complex neural network architectures to learn process data features and utilizes growing computational resources for efficient process optimization and control [37]. Feature extractor models like RBM could also be computationally intractable to train through classical training algorithms. Therefore, there arises a need to develop high-performance deep learning models for fault detection and diagnosis capable of overcoming limitations of the current machine learning paradigms carried out on state-of-the-art classical computers.

There are several research challenges towards developing QC-based process monitoring techniques that utilize deep learning architectures and ensure effective fault detection and diagnosis performance. One such challenge is to design deep learning models and architectures that can extract faulty features from small datasets, since in most industrial applications large amounts of data for faulty operations are seldom available. A further challenge lies in training of such deep architectures as their complexity increases with the number of hyper-parameters. Faults must be detected and diagnosed quickly for safety concerns that implies the training process should be performed with reasonable computational costs. Limitations of the classical training algorithms for deep learning models and QC devices also pose a computational challenge. It is crucial to develop techniques that leverage both QC and classical computers to overcome such challenges.

In this work, we develop QC-based model and methods for fault detection and diagnosis of complex process systems that efficiently extract several levels of features for normal and faulty process operations using deep RBM-based architectures. For complex process systems with high number of process measurements, training the RBMs is computationally challenging and might also result in suboptimal hyper-parameters that further affect the classification accuracy of fault detection models. To this end, we train the RBM-based network in the QC-based deep learning model with a quantum assisted training algorithm to overcome such computational challenges. The proposed model effectively detects faults in complex process systems by leveraging the superior feature extraction and deep learning techniques to facilitate proper discrimination between normal and faulty process states. Complexities such as nonlinearities between process variables and correlations between historical data can also be handled by this QC-based fault diagnosis model. The applicability of this QC-based deep learning method is demonstrated through two case studies on statistical process monitoring of the closed-loop continuous stirred tank reactor (CSTR) and the Tennessee Eastman (TE) process, respectively. These two processes are commonly used in benchmarking applications to measure and compare the performance of the fault diagnosis models. The CSTR simulation deals with a first-order reaction carried out in a tank with seven process variables recorded at each step that has three types of simulated faults, while the TE process is a relatively large industrial chemical manufacturing process with 52 process variables and 20 faults. Computational challenges stemming from the large size of the RBM used for the case studies are effectively tackled by the proposed QC-assisted training process. The obtained computational results for detecting anomalies are compared against state-of-the-art data-driven models and deep fault detection models trained on classical computers.

The major contributions of this work are summarized below:
- A novel QC-based deep learning model for detection and diagnosis of faults in complex process systems is proposed;
- The feature extractor network in the QC-based fault diagnosis model is trained with a novel training process that performs generative training assisted by quantum sampling;
- Case studies on CSTR and TE process are presented with comprehensive comparison against state-of-the-art fault detection methods using classical computers.

We first provide a brief background on RBMs and adiabatic quantum computing. These preliminaries in Section 2 are paramount to implementation of the proposed models and methods in this paper. The proposed QC-based deep learning model for fault diagnosis and quantum assisted methods are presented in Section 3. Two industrial case studies are presented in Sections 4 and 5 to demonstrate the effectiveness of the proposed model. These are followed by a discussion on quantum advantage perceived in the respective case studies in Section 6. Conclusions are drawn in the Section 7.

## 2. Background

### 2.1. Adiabatic Quantum Computing

An important architecture of quantum computing is the computational model of adiabatic quantum computing (AQC) that started out as an approach to solving optimization problems [38]. AQC permits quantum tunneling to explore low-cost solutions and ultimately yields a global minimum [39]. It also exhibits convergence to the optimal or ground state with larger probability than simulated annealing [39]. AQC devices intrinsically realize quantum annealing algorithms to solve combinatorial optimization problems giving birth to the paradigm of adiabatic quantum optimization (AQO). AQO is an elegant approach that helps escape local minima and overcomes barriers by tunneling through them rather than stochastically overcoming them as shown in Figure 1a. AQO can also be referred to as the class of procedures for solving optimization problems using a quantum computer.

In AQC, the computation proceeds by moving from a low-energy eigenstate of the initial Hamiltonian to the ground state of the final Hamiltonian. A Hamiltonian mathematically describes the physical system in terms of its energies, and corresponds to the objective function of an optimization problem in the final Hamiltonian [40]. The adiabatic optimization process evolves the quantum state towards a user-defined final problem Hamiltonian, while simultaneously reducing the influence of initial Hamiltonian in an adiabatic manner [39]. Tunneling between various classical states or the eigenstates of the problem Hamiltonian is governed by the amplitude of the initial Hamiltonian. Decreasing this amplitude from a very large value to zero drives the system into the ground state of the problem Hamiltonian that corresponds to the optimal solution of the objective function.

In order to solve optimization problems with AQC, they need to be formulated as an Ising model or quadratic unconstrained binary optimization (QUBO) problems. Such QC devices that are designed to implement AQO are commercially made available by D-Wave systems. The quantum processing unit on D-Wave devices is represented as a lattice of qubits interconnected in a design known as *Chimera* graph. Figure 1b is a subgraph of the *Chimera* lattice pattern that is typical of the D-Wave systems and their operation. The objective function represented as an Ising model or a QUBO problem has to be mapped to the qubits and couplers of the *Chimera* lattice. Mapping of variables to the qubits requires a process called minor embedding. Embedding is an important step since the *Chimera* lattice is not fully connected [41, 42]. The adiabatic optimization process follows after the mapping of the objective function onto the physical quantum processing unit that searches for low-energy solutions of the corresponding problem Hamiltonian [43]. The embedding and annealing schedule dictate the probability of recovering global optimal solutions .

The behavior of AQC systems in the presence of noise highly influences its performance and has been a subject of interest among researchers. Generic results for the Hamiltonian-based algorithm perturbed by particular forms of noise have also been reported [44]. Adiabatic computation requires the gap between the excited states and the ground states to be not too small. Adiabatic evolution is particularly susceptible to noise if this gap is small [45]. It has also been shown that under certain conditions, thermal interactions with environment can improve the performance of AQC [46]. Apart from thermal fluctuations, several internal and external factors contribute to the noise in quantum systems. Qubits in such devices can be affected by the electronic control components and material impurities, which give rise to the external and internal sources of noise, respectively. In the context of optimization, noisy qubits deviate the state of the system from a global optimal solution to sub-optimal solution state. However, from a machine learning perspective, such noisy behavior and measurement uncertainty in quantum systems can be exploited to approximate sample distributions that could be used to model the distribution of data, as will be introduced in *Quantum Generative Training* section. Despite several quantum advantages offered by AQC, the role of current quantum technologies for process systems engineering is not well established. To provide a better sense of the capabilities and challenges of AQC approaches, the presented background highlights the theories and practicums of quantum computers through several references. Some of the major publications that construe basic concepts underlying AQC can be found in [38] and [39]. Readers interested in implementation of AQC-inspired methodologies can also refer to methodology focused articles like [13] and [47].

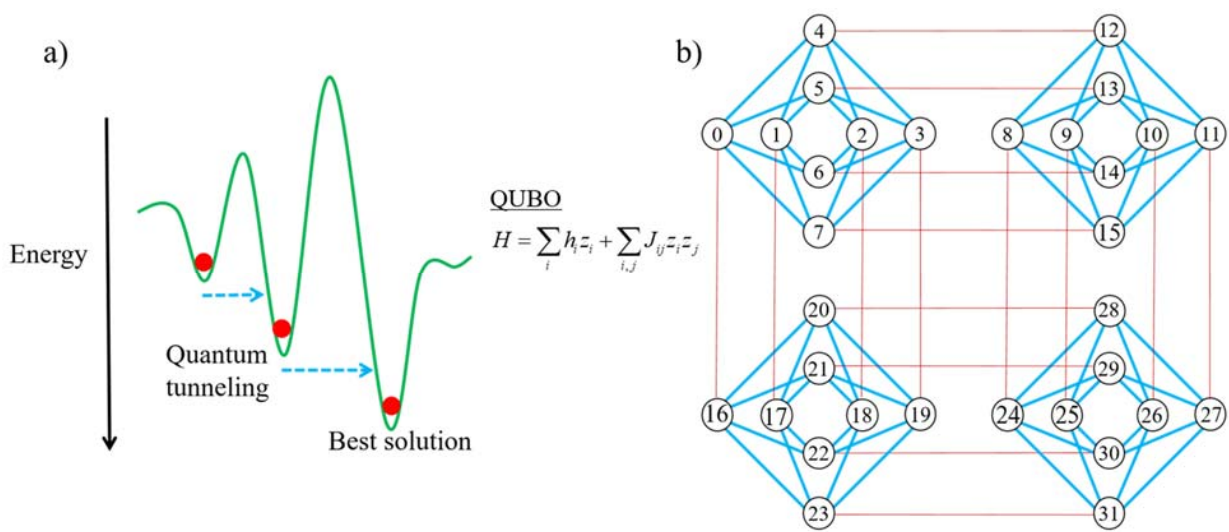

*Figure 1. a) Adiabatic quantum optimization (AQO) and b) Chimera architecture of the D-wave processing unit*

## 2.2. Restricted Boltzmann Machine

RBMs also termed as harmoniums [48] are interpreted as generative stochastic forms for artificial neural networks used to learn the underlying data distributions. In recent years, RBMs have been widely applied for pattern analysis and generation with applications in image generation [49], collaborative filtering for movie recommendations [50], phone recognition [51], and many more. As the name suggests, RBM is a restricted variant of Boltzmann machine that forms an undirected bipartite graph as shown in Figure 2, between neurons from two groups commonly termed as visible and hidden units. A RBM network with $m$ visible neurons and $n$ hidden neurons represent the observable data and the dependencies between the observed variables, respectively [52]. The hyper-parameters for this undirected bipartite graph are the weights and biases. For a pair of visible unit $v_i$ and a hidden unit $h_j$, a real valued weight $w_{ij}$ is associated with the edge between them. A bias term $b_i$ and $c_j$ are associated with the $i$th visible unit and $j$th hidden unit, respectively.

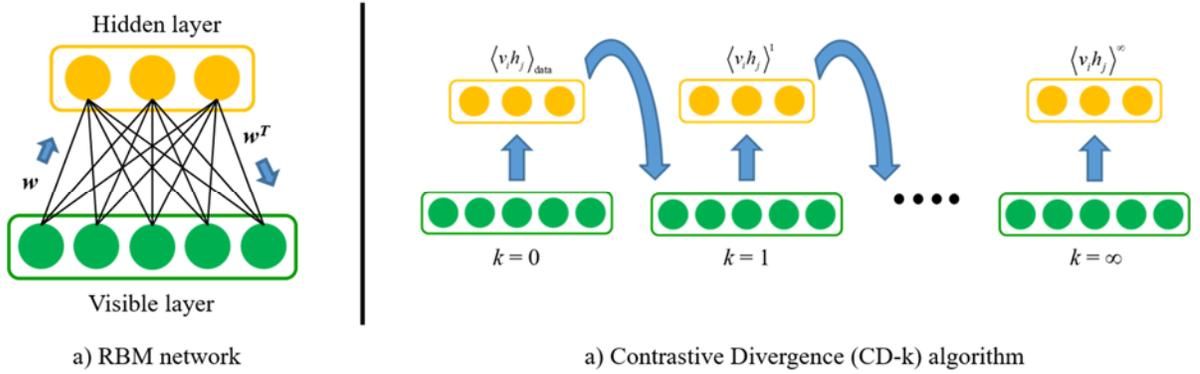

a) RBM network      a) Contrastive Divergence (CD-k) algorithm

*Figure 2. a) Schematic network of RBM and b) contrastive divergence algorithm*

The energy function of a RBM [53] for the joint configuration of binary or Bernoulli visible and hidden units $(\boldsymbol{v}, \boldsymbol{h}) \in \{0,1\}^{m+n}$ is given by $E(\boldsymbol{v},\boldsymbol{h})$ as shown in Eq. (1). Due to the absence of connections between units of the same layer, the state of the hidden variables is independent of the state of the visible variables and vice versa. A probability is assigned by the network to each possible pair of visible and hidden units through the RBM energy function as shown in Eq. (2), where the normalization constant or the partition function $Z$ is defined by summing over all possible pairs of visible and hidden vectors. This joint probability distribution is defined by a Gibbs or a Boltzmann distribution. Due to the conditional independence between the variables in the same layer, the conditional distributions factorize nicely and simple expressions for the marginal distributions of visible variables can be obtained. Eq. (3) gives the probability assigned to a visible vector $\boldsymbol{v}$ obtained by summing over all possible hidden vectors.

$$E(\mathbf{v},\mathbf{h}) = -\sum_{i \in \text{visible}} b_i v_i - \sum_{j \in \text{hidden}} c_j h_j - \sum_{i,j} w_{ij} v_i h_j \qquad (1)$$

$$p(\mathbf{v},\mathbf{h}) = \frac{1}{Z}\exp(-E(\mathbf{v},\mathbf{h})), \quad Z = \sum_{\mathbf{v'},\mathbf{h'}} \exp(-E(\mathbf{v'},\mathbf{h'})) \qquad (2)$$

$$p(\mathbf{v}) = \frac{1}{Z}\sum_{\mathbf{h}} \exp(-E(\mathbf{v},\mathbf{h})) \tag{3}$$

Generative training deals with determining the weights and biases that maximize the likelihood or log-likelihood of the observed data. To maximize the probability $p(\mathbf{v})$ assigned to the training data vector $\mathbf{v}$ by the RBM, the weights and biases of the network are updated such that the energy of the training data vector is lowered, while simultaneously raising the energy of the other training data vectors. The gradients of the log-likelihood of the training data with respect to the hyper-parameters of the RBM can be calculated from Eq. (4). The gradients can be interpreted as the difference between the expectation values under the distributions of training data and the underlying model.

$$\frac{\partial \log p(\mathbf{v})}{\partial w_{ij}} = \langle v_i h_j \rangle_{data} - \langle v_i h_j \rangle_{model} \tag{4}$$

Learning rules to update the values of weights and biases can be derived from these log-likelihood gradients in order to maximize the log probability with stochastic gradient ascent. Eqs. (5), (6), (7) describe the update rules where $\varepsilon$ is the learning rate and $\alpha$ is the momentum. The terms $\langle v_i h_j \rangle_{data}, \langle v_i \rangle_{data}, \langle h_j \rangle_{data}$ are the clamped expectation values with a fixed $\mathbf{v}$ and can be efficiently computed from training data using Eq. (8). This equation provides an unbiased sample of the clamped expectations where $\sigma(x)$ is the logistic sigmoid function defined by $\sigma(x) = 1/(1 + e^{-x})$. Eq. (9) also produces unbiased samples of visible states, give a hidden vector $\mathbf{h}$.

$$w_{ij}^{t+1} = \alpha w_{ij}^t + \varepsilon \left( \langle v_i h_j \rangle_{data} - \langle v_i h_j \rangle_{model} \right) \tag{5}$$

$$b_i^{t+1} = \alpha b_i^t + \varepsilon \left( \langle v_i \rangle_{data} - \langle v_i \rangle_{model} \right) \tag{6}$$

$$c_j^{t+1} = \alpha c_j^t + \varepsilon \left( \langle h_j \rangle_{data} - \langle h_j \rangle_{model} \right) \tag{7}$$

$$P(h_j = 1 | \mathbf{v}) = \sigma \left( c_j + \sum_{i \in \mathbf{v}} w_{ij} v_i \right) \tag{8}$$

$$P(v_i = 1 | \mathbf{h}) = \sigma \left( b_i + \sum_{j \in \mathbf{h}} w_{ij} h_j \right) \tag{9}$$

The model expectations $\langle v_i h_j \rangle_{model}, \langle v_i \rangle_{model}, \langle h_j \rangle_{model}$ are difficult to estimate. They can be computed by randomly initializing the visible states and performing Gibbs sampling for a long time. However, this can be computationally intractable as the number of visible and hidden units increases [54]. Hinton proposed a faster learning algorithm called contrastive divergence (CD) learning [55] that has become a standard way to train RBMs. Rather than approximating the model expectations by running a Markov chain until equilibrium is achieved, the $k$-step CD learning (CD-k) runs the Gibbs chain for only $k$ steps to yield the samples $\langle v_i h_j \rangle^k, \langle v_i \rangle^k, \langle h_j \rangle^k$ as shown in Figure 2b. This learning algorithm works well despite the *k*-step reconstruction of the training data crudely approximating the model expectations [55]. Theoretically, as $k \to \infty$ the update rules converge to the true gradient. However, in practice the updates are computed using a single-step (*k*=1) reconstruction to achieve good enough performance.

Many significant applications use real-valued data nowadays for which the binary RBM would produce poor logistic representations. In such cases, a modified variation of the RBM can be used by replacing Bernoulli visible units with Gaussian visible units [54]. The energy function then takes the form of Eq. (10), where $\sigma_i$ is the standard deviation of the Gaussian noise for the *i*th visible unit. CD-1 can be used to learn the variance of the noise, but it is much more complicated than the binary case. An easier alternative is to normalize each data component to have zero mean and unit variance, and then use noise-free models. The variance $\sigma^2$ would be unity in this case.

$$E(\mathbf{v},\mathbf{h}) = -\sum_{i \in \text{visible}} \frac{(v_i - b_i)^2}{2\sigma_i^2} - \sum_{j \in \text{hidden}} c_j h_j - \sum_{i,j} \frac{v_i}{\sigma_i} h_j w_{ij} \qquad (10)$$

Deep architectures can be constructed by stacking layers of RBMs together. Such deep architectures are termed as deep belief networks (DBNs) where each RBM sub-network's hidden layer serves as the visible layer for the following RBM layer [56]. DBNs are trained in a greedy fashion by sequentially training each RBM layer. There have been many implementations and uses of DBNs in real-world applications due to their versatility and effective multiple-level feature extraction capabilities [57].

## 3. Quantum Computing-based Fault Diagnosis Model

The proposed QC-based deep learning model utilizes a two-step strategy, namely quantum generative training followed by supervised discriminative training using class labels. The first step involves using two DBN sub-networks to extract features from historical process data. Features at different levels are extracted for normal state along with each of the faulty state through the quantum generative training process. The DBN sub-networks, namely DBN-N and DBN-F are trained separately using normal and faulty training datasets, respectively. Using two DBN sub-networks is accounted for by the variability of the extracted features from each sub-network. Several computational experiments performed with only a single DBN also reveal that the extracted features are affected by the relatively inferior generalized feature extraction capabilities of energy-based models like DBN. The amount of training data required to achieve maximum performance depends both on the model complexity and the complexity of training algorithm. Ten times more data samples than the number of input dimensions can be used as a statistical heuristic [58]. The analysis of dataset size versus model skill is termed as learning curve and can also be conducted to obtain bounds on the size of training dataset for a required precision of performance measurement. The input to the fault diagnosis model is a data vector with *d* dimensions that correspond to each process variable. In order to classify the state of this data vector, outputs from the pre-trained sub-networks DBN-N and DBN-F that serve as $k$ dimensional approximations of the input data, are combined together.

The second step uses the combined approximate $2k$ dimensional vector. It is passed on to the local classification sub-network that predicts the state of the original input data vector. The local classification deep neural network-based architecture yields the probabilities of two possible states, normal and faulty. The local classifier follows a supervised discriminative learning strategy that uses class labels as an extra output layer. A graphical representation of the proposed QC-based fault diagnosis model is shown in Figure 3. Since the performance of DBN-based networks is known to be sub-optimal due to the presence of several local minima, generative training helps locating a desired local neighborhood near a good optimum while discriminative training further refines the optimum by fine-tuning the model parameters.

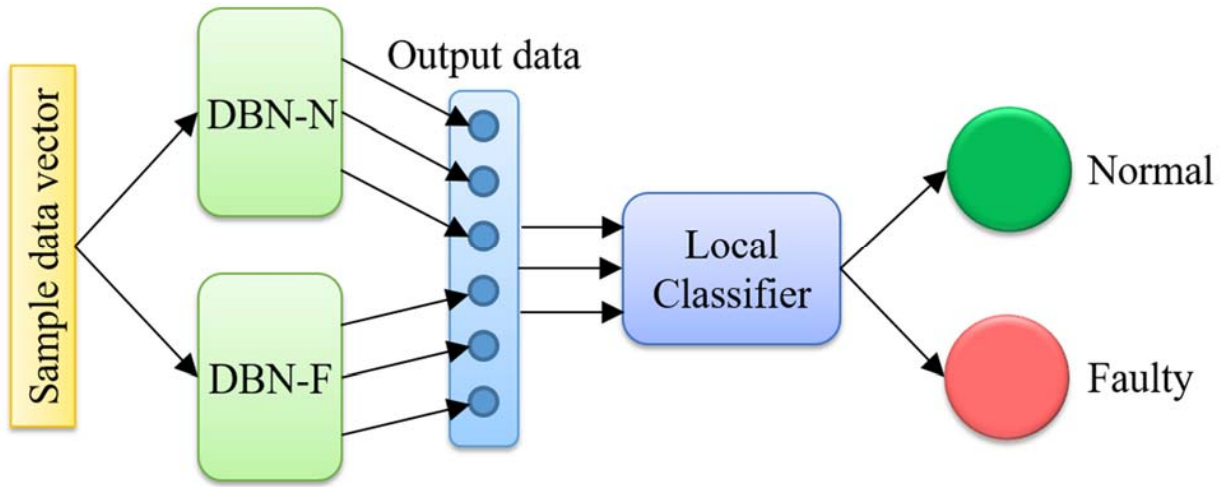

*Figure 3. Repeating sub-network in the proposed QC-based fault diagnosis model that uses deep belief networks and local classifier to predict the state of the data samples*

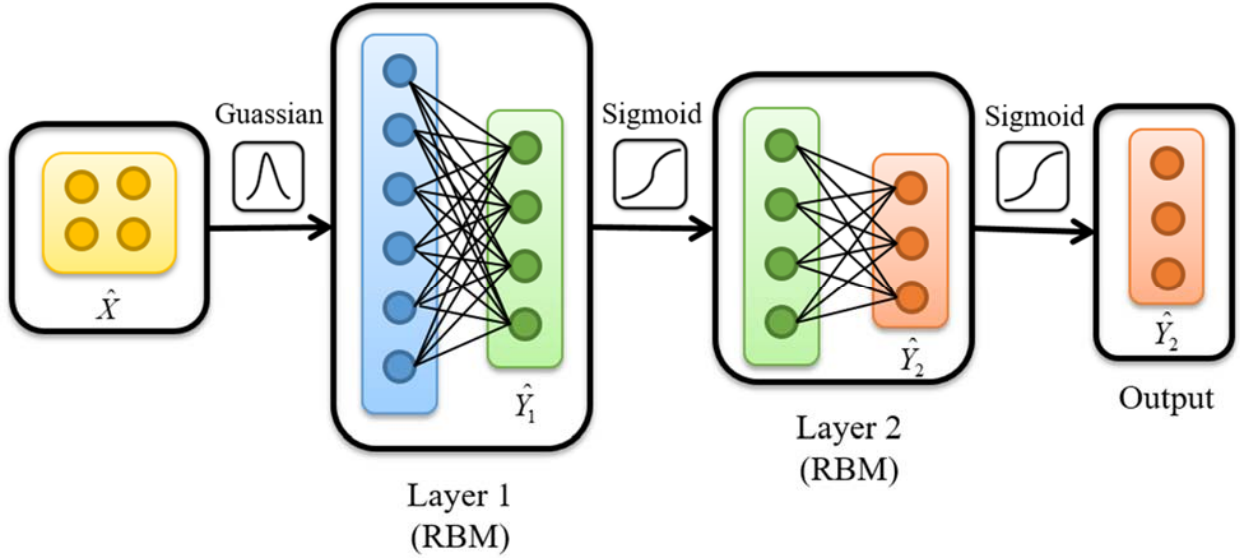

*Figure 4. Deep belief network architecture used in the repeating sub-network of the QC-based fault diagnosis model that produces a high level abstraction of the input data*

## 3.1. Quantum Generative Training

As mentioned in the previous subsection, two DBN sub-networks DBN-N and DBN-F extract the underlying features for normal and faulty process states through the quantum generative training process. Each DBN sub-network comprises of two RBMs represented in Figure 4 that are stacked atop each other and trained sequentially. The RBM must be extended to handle continuous valued inputs as most complex process systems provide continuous real-valued data. Therefore, the first RBM layer uses $d$ Gaussian visible units with $m$ Bernoulli or binary-valued hidden units. Input to this layer is the historical process data vector $\hat{X}_{n \times d}$ with $n$ samples and $d$ process data dimensions. The model parameters for this RBM layer are denoted as $W_g, B_g$, and $C_g$. $W_g \in \mathbb{R}^{d \times m}$ is the connection weights matrix between the visible and hidden nodes of the RBM, while $B_g \in \mathbb{R}^d$ and $C_g \in \mathbb{R}^m$ are the visible and hidden bias vectors, respectively. This RBM layer is trained by the CD-1 algorithm with an appropriate learning rate that prevents the RBM from under-fitting or over-fitting the historical data. The weights and biases for this layer are updated such that the reconstruction loss between the input data vector $\hat{X}$ and the reconstructed data vector $\hat{X}_r$ is minimized. The output from the first RBM layer $\hat{Y}_1 \in \{0,1\}^{n \times m}$ is generated by multiplying the input data vector with the weights matrix and adding the corresponding hidden biases followed by a sigmoid activation function operation given in Eq. (11).

$$\hat{Y}_1 = \sigma\left(\hat{X} \cdot W_g + C_g\right) \quad (11)$$

Following the first RBM layer, the second RBM layer in the DBN extracts higher level features from the process data. Deep network architectures are always preferred over shallow networks, but increasing model complexity requires large amount of training data to achieve optimum model skill. Also, increasing the number of layers introduces size constraints on the following layers and might limit the model performance. Computational experiments conducted with one RBM layer yield a lower performance than relatively deeper architectures. Therefore, two RBM layers are used in the DBN sub-networks. Binary output vector $\hat{Y}_1$ obtained from the first RBM layer serves as input to this RBM layer. Therefore, the visible and hidden units of the second RBM are modeled as Bernoulli units. The weights matrix $W_b \in \mathbb{R}^{m \times k}$, visible bias $B_b \in \mathbb{R}^m$, and hidden bias $C_b \in \mathbb{R}^k$ form the model parameters for this layer that need to be optimized. The update rules for these model parameters require the computation of model expectations $\langle v_i h_j \rangle_{model}$, $\langle v_i \rangle_{model}$, and $\langle h_j \rangle_{model}$. Since the CD algorithm approximates the gradient for the update rules with a larger variance that might not always lead to the maximum likelihood estimate of the model parameters, the model expectations are estimated using quantum sampling implemented through a quantum computer.

The AQC devices are explicitly built for optimization purposes by determining the ground state of the problem Hamiltonian. However, there have been experimental evidence suggesting that under certain conditions such devices sample approximately from a Boltzmann distribution at an effective temperature [59, 60]. The final states of the qubits are effectively described by a Boltzmann distribution when the strengths of the fields and couplings on the device are sufficiently small. Due to the presence of non-ideal interactions between the qubits and the environment, the AQC device can be used as a sampling engine [17]. A natural resemblance exists between the problem Hamiltonian taking the form of a QUBO problem and the energy function of the RBM with Bernoulli units. Quantum sampling exploits this by embedding the RBM energy function onto the AQC device. The distribution of the excited states of the qubits can then be modeled as a Boltzmann distribution given in Eq. (12). An unknown scale parameter $\beta_{eff}$ dictates the effective temperature at which samples are drawn from the underlying Boltzmann distribution. The value of this parameter depends on the operating conditions of the AQC device, and it is a direct link between the problem Hamiltonian and the energy function. Although some techniques have been proposed that estimate the effective temperature [61], a constant value for $\beta_{eff}$ is empirically selected depending on the size of the RBM. Samples drawn from an AQC device follow a trend as shown in Figure 5.

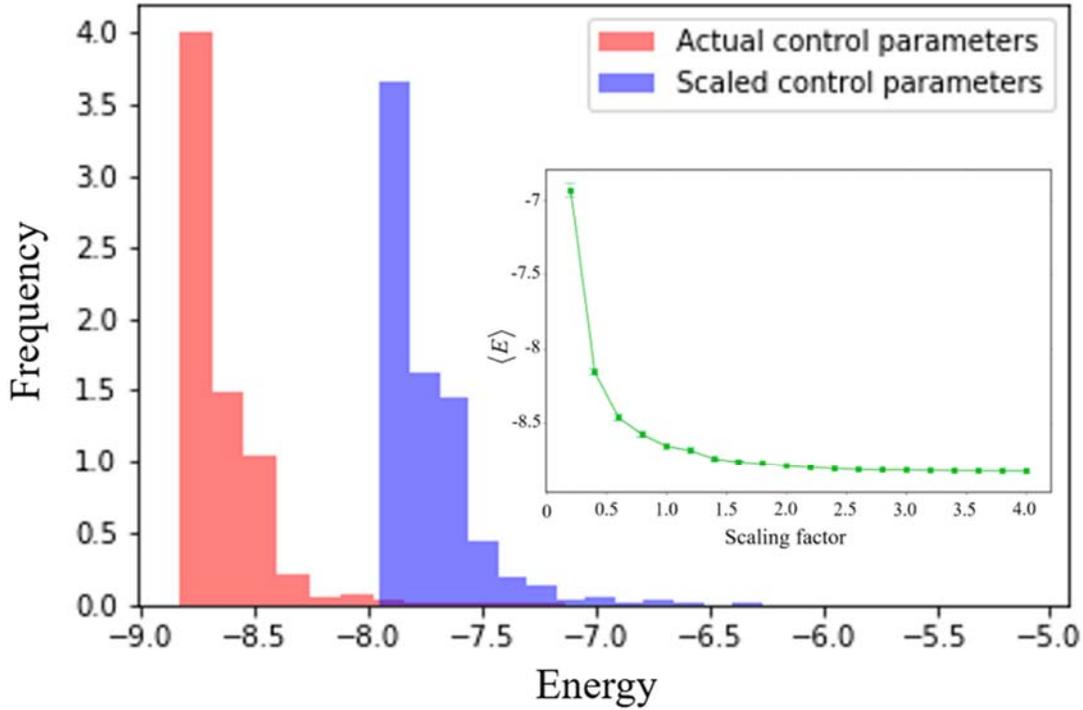

*Figure 5. RBM Energy histogram obtained for two sets of control parameters obtained by increasing the actual parameters by a scaling factor along with the effect of scaling factor on the average energy*

Control parameters used for the quantum sampling process are equivalent to the weights and biases of the RBM energy function provided that the scale parameter is unity. $\beta_{eff}$ can also be estimated by adjusting the actual control parameters by a user-defined scaling factor and analyzing the difference between the histogram of samples drawn from an AQC device as shown in Figure 5. Selecting an appropriate scaling factor is a crucial task; increasing the scaling factor tends to reduce the average energy of the samples drawn through quantum sampling. Setting the value of the unknown scale parameter to one eliminates the need for analytically calculating $\beta_{eff}$ at each iteration of the training process, and sparingly reduces the required computational resources and time. It should also be noted that the AQC device could be used over the cloud, thus providing a cost-effective alternative over buying an expensive AQC device.

$$P(v,h) = \frac{1}{Z}\exp\left(-\beta_{eff} E_{RBM}\right) \qquad (12)$$

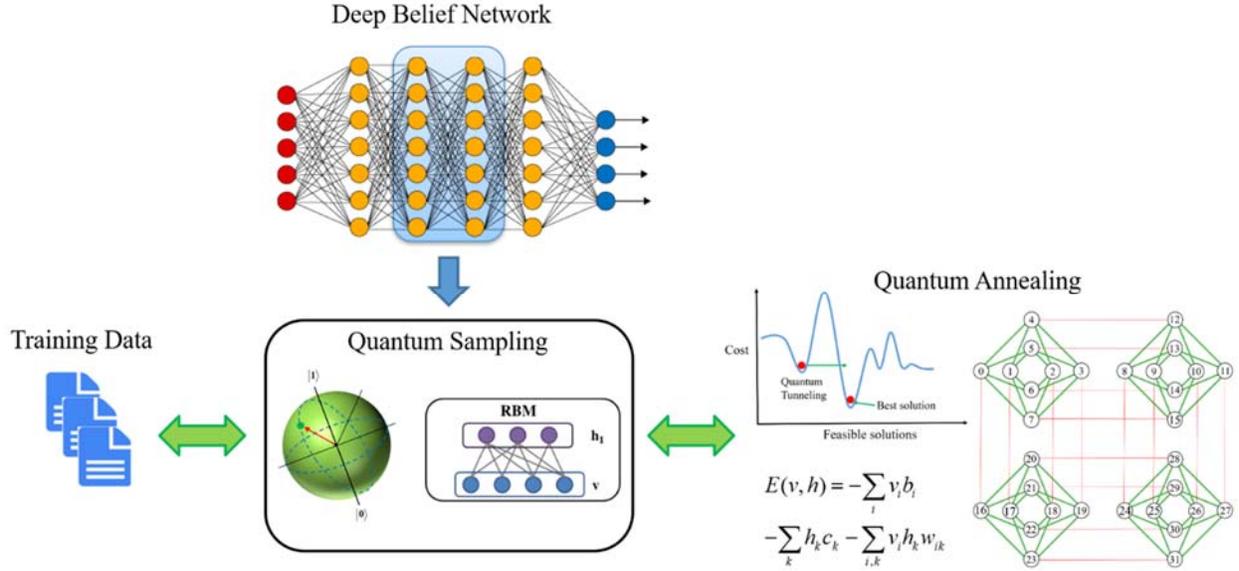

*Figure 6. Quantum generative training through quantum sampling*

With the approximate knowledge of the underlying Boltzmann distribution, the model expectations are computed by drawing several samples corresponding to the RBM energy function by quantum sampling. Eqs. (13), (14), (15) use $N$ samples drawn from adiabatic optimization runs to calculate the corresponding model expectation values required to update the model parameters. Figure 6 summarizes the quantum generative training process that uses quantum sampling to find the maximum likelihood estimates of the corresponding model parameters.

$$\langle v_i h_j \rangle_{model} = \frac{1}{N} \sum_{n=1}^{N} v_i^n h_j^n \tag{13}$$

$$\langle v_i \rangle_{model} = \frac{1}{N} \sum_{n=1}^{N} v_i^n \tag{14}$$

$$\langle h_j \rangle_{model} = \frac{1}{N} \sum_{n=1}^{N} h_j^n \tag{15}$$

The update rules for the weights and biases of the second RBM in the DBN sub-network given in Eq. (5) converge to the minimum cross-entropy loss between the original input and the reconstructed input vector. The output from the second RBM layer $\hat{Y}_2 \in \mathbb{R}^{n \times k}$ bounded by [0,1] is obtained by multiplying input data vector with the weights matrix and adding the corresponding hidden biases followed by a sigmoid activation function operation given in Eq. (16). Output of the generative training model $\hat{Y}_2$ is a transformed version of the original input data vector $\hat{X}$. This transformation can be considered as a higher-level abstraction of the historical process data and can be used as an input to the corresponding classifier to determine the state of the input data sample in the QC-based fault diagnosis model.

$$\hat{Y}_2 = \sigma\left(\hat{Y}_1 \cdot W_b + C_b\right) \tag{16}$$

## 3.2. Discriminative training

High level abstraction of the process data generated by the pre-trained DBN-N and DBN-F sub-networks are concatenated together and is passed as input to the fully connected network. DBN-N produces normal state abstractions while the DBN-F sub-network produces faulty state abstractions. The architecture of the local classifier sub-network is shown in Figure 7. Using class labels for normal and faulty data, discriminative training is performed in a supervised manner. This is accomplished by adding a fully connected network with a single hidden layer followed by a soft-max layer that predicts the probabilities of normal and faulty states.

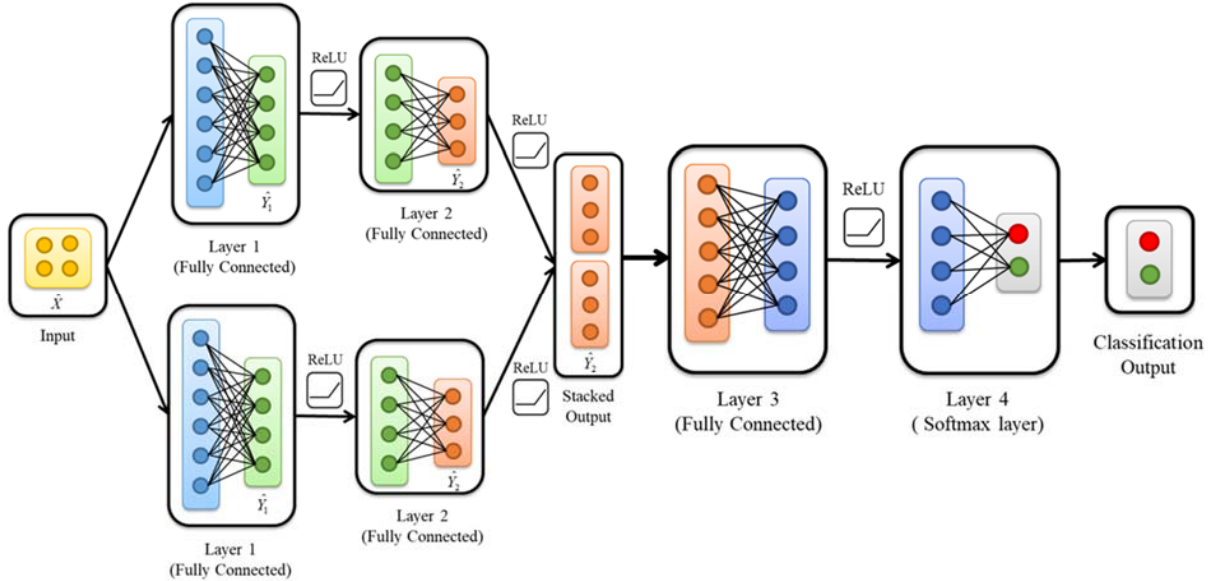

Figure 7. Local classifier architecture that identifies normal or faulty data samples

The weights matrix $W_f \in \mathbb{R}^{2k \times 2k}$ and bias vector $b_f \in \mathbb{R}^{2k}$ form the model parameters for the fully connected layer that connects each input to every hidden neuron. Nonlinear combinations of the extracted features can be easily learned with a fully connected layer which is a major component of the discriminative training process. The output generated by this layer $\hat{Y}_3 \in \mathbb{R}^{n \times 2k}$ is used to predict the score of normal or faulty class; it is obtained by summing the bias and product of weights matrix with the input vector as shown in Eq. (17) followed by a ReLU activation function operation. As the process data can be in either of the two states, normal or faulty, weights vector $W_s \in \mathbb{R}^{2k \times 2}$ and bias vector $b_s \in \mathbb{R}^2$ predict the final class scores using the soft-max activation function in Eq. (18). Model parameters for the DBN-based sub-networks are fine-tuned by retraining the local classifier neural network classically with the backpropagation algorithm that performs supervised learning of neural networks using gradient descent. The gradients of the loss function are estimated with respect to the model parameters of the local classifier sub-network, in order to iteratively update the model parameter values. Minimizing the categorical cross-entropy loss for the classifier yields maximum likelihood estimates of the model parameters. The cross-entropy loss can be computed with the predicted class scores as shown in Eq. (19), where $Y^T$ are the true fault labels.

$$\hat{Y}_3 = \text{ReLU}\left(\hat{Y}_2 \cdot W_f + b_f\right) \tag{17}$$

$$P_i = \frac{\exp\left(\hat{Y}_3 \cdot W_s^i + b_s^i\right)}{\sum_i \exp\left(\hat{Y}_3 \cdot W_s^i + b_s^i\right)} \tag{18}$$

$$Loss_{CE} = -\sum_i Y_i^T \cdot \log P_i \tag{19}$$

A QC-based fault diagnosis model for individual process faults is obtained by following the quantum generative training and discriminative training process. The DBN-N sub-network in the generative model is trained only once and can be re-used for each diagnosis model. To detect the unknown state of the process data sample, both normal state and faulty state abstractions of the data sample generated as the output of the DBN-based generative model are merged. The local classifier then predicts the probabilities that the data sample belongs to normal or faulty states. A threshold probability of 0.5 further detects the state of the new process data sample.

## 4. Application: Continuous Stirred Tank Reactor

A closed-loop feedback controlled CSTR is used to evaluate the performance of the proposed process monitoring method. The CSTR simulation continuously carries out a first-order exothermic reaction in a jacketed tank with constant holdup and records normal and faulty data at specific intervals [62]. Figure 8 shows the schematic of the CSTR case study. The concentrations, temperatures, and volumetric flow rate account for the seven process variables in the CSTR process simulation. Three faults are investigated in this case study, where the faults are caused due to errors in the reactor temperature measurement, decay in catalyst activity and fouling in the cooling jacket. Fault 1 is a sensor fault caused by a drift in the readings of reactor temperature. Fault 2 is simulated by introducing a slow decay in catalyst activity, while fault 3 is a fouling fault in the cooling jacket of the CSTR. Normal and faulty datasets with 1200 samples are recorded at a sampling interval of one minute for training the proposed QC-based fault diagnosis model. In faulty datasets, the fault is introduced after 200 minutes of normal operation. A testing set with 600 samples recorded for 10 hours is used to validate the trained model for both normal and faulty states, where the fault is introduced after 100 minutes of normal operation. The training and testing datasets corresponding to both normal and faulty states used for this case study are provided in the Supplementary Information.

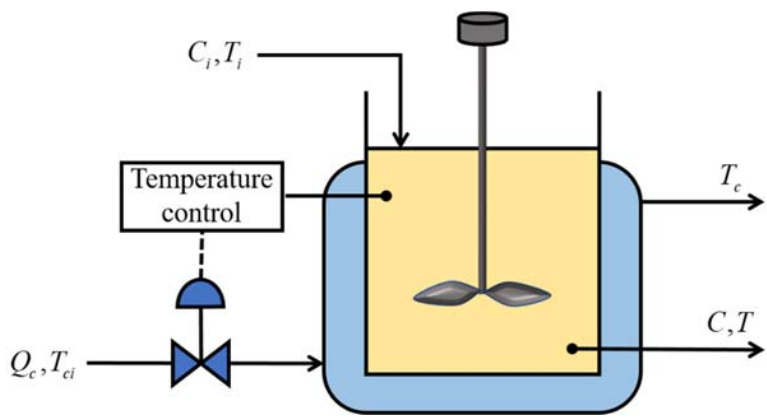

*Figure 8. Schematic of the continuous stirred tank reactor (CSTR)*

## 4.1. Experimental Settings

As the process data has strong temporal correlations, its dynamic characteristics must be considered. The data pre-processing step is clearly shown in Figure 9 and can be described as follows. Assuming a $d$ dimensional raw dataset recorded for $T$ time-steps, the length of time window $N$ is a fundamental element of the pre-processing step. The process data matrix of size $N \times d$ is expanded into a one dimensional vector and accounts for a single sample in the input data vector. A similar procedure is repeated for the process data labels in order to preserve the size and information corresponding to the input data. This pre-processing is the first step towards dealing with temporal autocorrelations between the process data variables with the size of time window $N$ determined empirically by performing several computational experiments. A RBM with Gaussian visible units interacts with this input data, because the recorded process data are continuous and real-valued. In order to eliminate the complexities encountered during training a RBM with Gaussian visible units, the input data corresponding to normal and three faulty states is normalized to have zero mean and unit variance.

The size of dynamic time window $N$ is set to 4 which implies 28 visible Gaussian units in the first RBM layer of the DBN-based sub-network. This layer consists of 15 Bernoulli hidden units and produces output by applying perceptron operation to the input without sampling from a Gaussian distribution. This output is used as input data vector for the following RBM layer with 15 visible and 8 hidden units. The second RBM layer uses Bernoulli units and generates an output by performing a sigmoid operation on the corresponding perceptron output. Learning rate of 0.001 is used to train the RBM layer with Gaussian visible units and 0.01 for the layer with binary visible and hidden units. The momentum for weight and bias updates is set to unity. Both sub-networks DBN-N and DBN-F follow exactly the same architecture and are trained through the quantum generative training process. Cross-entropy loss is used as a performance metric to track the training progress. Mean square loss can also be used as a viable substitute for cross-entropy loss. The hidden layers or the outputs in the DBN-N and DBN-F sub-networks are merged together as a single layer with 16 neurons. This ensures that the higher level abstractions of the normal and faulty states are processed together. A fully connected layer with 16 neurons followed by a soft-max layer is attached to the merged outputs and forms the basis of the discriminative training. With the weights and biases obtained through the quantum generative training as starting points, the complete network is retrained with Adam optimizer to minimize the categorical cross-entropy loss.

In order to draw samples from the AQC-based device, D-wave's 2000Q quantum processing unit is used remotely over the cloud. The model expectations required to compute the weight and bias updates are calculated with these samples. This AQC-based device uses 2,048 qubits and 5,600 couplers that limit the size of fully connected RBM energy function with an equal number of visible and hidden units to 52 units in each layer. 1,000 anneal runs are performed with each run lasting for 20μs on this quantum processing unit. An embedding scheme for the corresponding RBM energy function is determined in a heuristic manner and the obtained graph minor is re-used to eliminate unnecessary complications with the effective temperature parameter $\beta_{eff}$. This parameter is set to a constant value of one for the CSTR case study.

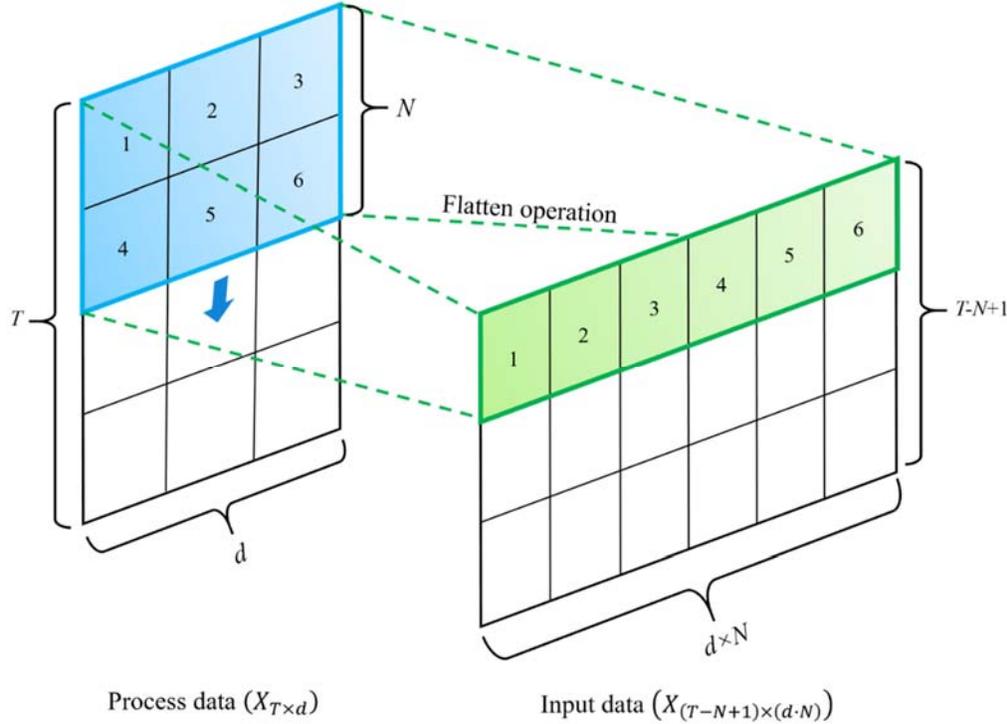

*Figure 9. Data pre-processing step for the CSTR case study*

## 4.2. Fault Detection

Several computational experiments are conducted with the aforementioned experimental settings to demonstrate the viability of the proposed QC-based fault diagnosis model. The local classifier detects the state of each sample and classifies it as normal or faulty by predicting the likelihood of individual states. It is important to note that the DBN-F sub-network is trained using the entire faulty dataset consisting of data samples corresponding to each individual fault. This allows the local classifier to effectively learn a discernible pattern between normal and faulty data samples. As a result, the proposed architecture is sensitive towards previously unseen states and is capable of detecting samples belonging to unknown faulty states. A probability control limit of 0.5 is used to classify the input data vector as normal or faulty. The fault detection rate (FDR) and the false alarm rate (FAR) are reported for each fault classified with the proposed QC-based fault diagnosis model in Table 1. FDR is defined as the fraction of faulty samples that are accurately detected, and FAR is the fraction of normal data samples that are incorrectly classified as faulty. If $p$ is the number of fault samples that are detected as faulty and $q$ is the number of normal samples detected as faulty, then FDR and FAR can be computed with Eqs. (20) and (21), respectively. FDRs for the CSTR case study estimated with canonical variate dissimilarity analysis (CVDA) [62] are also reported in Table 1, where the control limits for fault detection are computed with the $T^2$ statistic.

$$FDR = \frac{100 \cdot p}{\text{total count of faulty samples}}\% \quad (20)$$

$$FAR = \frac{100 \cdot q}{\text{total count of normal samples}}\% \quad (21)$$

*Table 1. Fault detection results of the local classifier in the proposed QC-based deep learning model for the CSTR case study*

| Fault | FDR (%) | |
|---|---|---|
| | CVDA | QC-based model |
| 1 | 43.83 | 100 |
| 2 | 64.19 | 82.83 |
| 3 | 71.71 | 74.42 |

It can be clearly seen that the FDR for the QC-based fault diagnosis model significantly improves for the sensor Fault 1. However, this may be accompanied by an increase in the number of false positives. As for the parametric faults 2 and 3, the FDRs are higher than or comparable to that of the detection rates obtained by CVDA. This improvement in detection rates can be attributed to the accurate capture of nonlinear process behavior by the QC-based fault diagnosis model. It is a well-known fact that the performance of deep architectures depends on the hyper-parameters like the number of neurons used. Therefore, we also generate heatmaps for FDRs as functions of the number of hidden units in the DBN-N and DBN-F sub-networks to determine the optimal configuration for the number of neurons in the hidden layers of the DBN sub-networks.

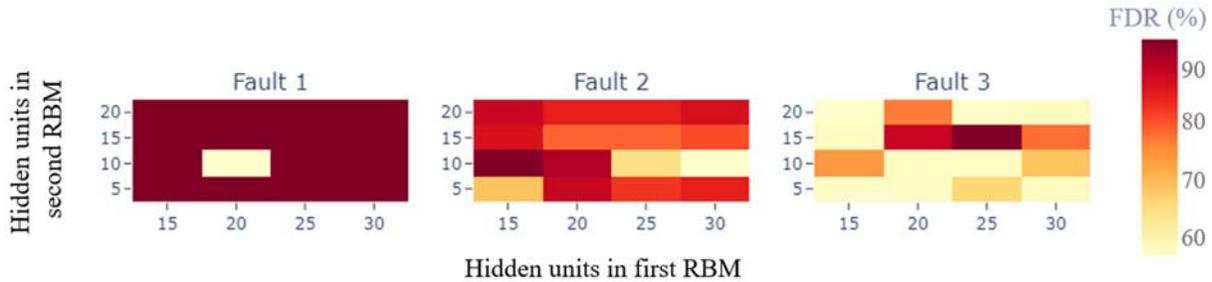

*Figure 10. FDR heatmaps of the DBN sub-networks for the CSTR case study*

The FDR maps for all three faults as shown in Figure 10 indicate that the detection rates for fault one are high for almost all DBN architectures. However, the detection results for fault two and fault three are relatively non-uniform. In case of fault two, for a fixed number of hidden units in the Bernoulli RBM, the FDRs gradually decrease with an increase in the hidden neurons in the first RBM layer. Alternatively, no discernible pattern is observed in the FDRs for fault three. The choice of best performing DBN architecture with 15 and 8 hidden units in the first RBM and the second RBM layer, respectively, can be clearly justified from these FDR heatmaps.

Among the 1,500 faulty samples in the testing dataset, the local classifier accurately classifies the dynamic input data samples with an average detection rate of 86.08%. This implies that the 13.92% of the faulty samples are missed. Compared to the missed detection rate of 40.42% in the CVDA technique, the QC-based fault diagnosis model clearly outperforms this fault monitoring technique. The average FAR rate obtained through performing multiple computational experiments for the CSTR case study is 19.41%. These false alarms can be accounted to the difficulty in differentiating between normal and faulty state one, as fault one deviates only slightly from normal operation and produces higher number of false alarms. Although the proposed QC-based fault diagnosis model may be accompanied by one false alarm for every five normal samples, it clearly outperforms the classical CVDA technique for all three faults with a significantly higher fault detection rate. Identifying faults for the CSTR case study is trivial and can be associated with the small-scale nature of the process system with lower number of process variables and fewer simulated faults. Performing fault identification for the CSTR case study manually by visual inspection of the recorded process variables could be much more cost efficient than implementing advanced fault diagnosis methodology for minor pattern recognition. In the next case study, automatic fault identification is essential due to the large-scale nature of the nonlinear process system for which we implement a QC-based deep learning model for fault identification.

## 5. Application: Tennessee Eastman Process

The TE process [63] is one of the popular benchmark problems for process monitoring, so it is used to test the proposed QC-based fault detection model in this section. It is a typical chemical process that produces two main products with five major process units, reactor, stripper, separator, compressor, and mixer. The TE chemical process simulation has 52 variables, containing 41 measured variables and 11 manipulated variables. In this TE process system, 20 fault states along with a normal state have been simulated, and process data has been recorded for each state. For each fault, 1,200 samples are recorded for 75 hours at a sampling interval of three minutes. Fault is introduced in the system after 10 hours of normal operation, meaning for each faulty dataset the first 200 samples correspond to normal process operation while the remaining 1,000 are faulty data samples. The dataset for normally operated process data is recorded for straight 48 hours without any disturbance. To validate the trained QC-based fault diagnosis model, a testing dataset with 600 samples is also recorded for 30 hours for normal and faulty states with the faults introduced after 10 hours of normal operation. Both training and testing datasets corresponding to normal and faulty states are provided in the Supplementary Information.

### 5.1. Experimental Settings

The recorded historical process data for the TE process are continuous and real-valued. Binary models might produce poor representations of such data, so an RBM with Gaussian visible units is the first interaction of the training process data with the proposed QC-based model. As mentioned earlier, the CD-1 algorithm might be inefficient in learning the variance of the Gaussian noise associated with each visible unit. In order to eliminate this variance, the training data corresponding to the normal state and the 20 faulty states are normalized to have zero mean and unit variance. Several fault detection methods implement variable selection pre-processing to consider the process variables with the highest influence on the data. However, for this case study all of the 52 process variables are used as input to the proposed model without eliminating any portion of the recorded dataset.

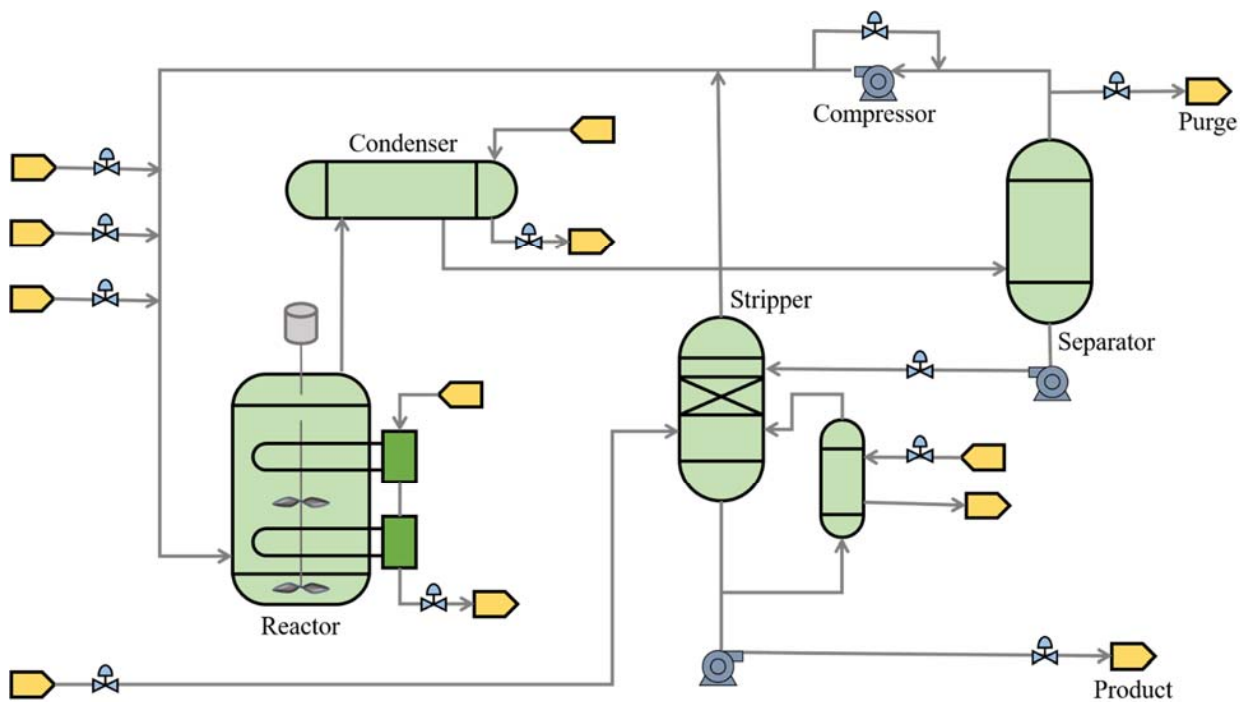

*Figure 11. Schematic of the Tennessee Eastman (TE) chemical process*

The sub-networks DBN-N and DBN-F use the same architectural configuration to produce abstractions of the normal and faulty states, respectively. The first RBM layer in the DBN-based sub-networks consists of 52 visible Gaussian units and 26 Bernoulli hidden units. This RBM layer is set up to produce an output by simple perceptron operation without sampling from the underlying Gaussian distribution. The following RBM layer used 26 visible units corresponding to the hidden layer of the first RBM layer along with 20 hidden units. Output obtained from the second RBM layer is produced by sampling from a binomial distribution with the hidden unit values as the means. A learning rate of 0.01 and momentum of one is used to train the DBN sub-networks via quantum generative training. Learning rate for RBM with Gaussian units should be at least one order of magnitude less than the corresponding binary RBM. Cross-entropy loss is used as a performance metric to track the progress of the quantum generative training process. A data vector with 40 dimensions is obtained after the high-level abstractions from the DBN-N and DBN-F sub-network are concatenated. Fully connected layer with 40 neurons is attached to this input and forms the major component of the discriminator sub-network. Fine-tuning of the weights and biases obtained through quantum generative training is performed by training the discriminator with the Adam optimizer to minimize the categorical cross-entropy loss.

The quantum generative training process draws samples from the AQC-based quantum computer for quantum sampling to approximate the model expectations. D-Wave 2000Q quantum processor with 2,048 qubits and 5,600 couplers is used remotely over the cloud for all computational experiments involving QC-based fault detection model. The anneal schedule runs for 20μs on this processor. To compute the model expectations, 1000 anneal reads are used implying the drawing of 1000 samples from the quantum computer. For a single RBM instance, an embedding scheme for the corresponding RBM energy function is found through a heuristic technique. Drawing samples from the quantum computer for the energy function requires the use of the same embedding scheme. It is important to re-use the same graph-minor in order to minimize the variation in the effective temperature dependent parameter $\beta_{eff}$. For this case study, the value of the unknown scale parameter is set to unity to avoid further complications associated with the hyper-parameter learning rules.

## 5.2. Fault Detection

The above experimental settings are used to conduct several computational experiments for all 20 faults. The output of the local classifier lies between 0 and 1 representing the likelihood of the data sample belonging to either normal or faulty state. A threshold probability of 0.5 is used to detect the state of the sample. The diagnosis results of the local classifier for each fault consist of FDRs and FARs computed for both the training and validation datasets and are given in Table 2. The FDRs and FARs are computed using Eqs. (20) and (21), respectively. It should also be noted that the DBN-F sub-network is trained using the entire faulty dataset consisting of data samples corresponding to each individual fault. This allows the local classifier to clearly detect abnormal data samples belonging to previously unknown faulty states.

Heatmaps of the FDRs determined by the local classifier in the QC-based fault diagnosis model are generated for each fault to perform a grid search for the best performing network architecture. Optimal neural network configuration is obtained by locating the maximum FDRs in these heatmaps. Figure 12 shows the FDR maps for the TE process as a function of the number of hidden units in the first and second RBM layer of the DBN-N and DBN-F sub-networks. Eight faults simulated in the TE process demonstrate a uniform FDR map irrespective of the size of the DBN architecture used. FDR maps for faults 1 and 18 are uniform with very few exceptions. Faults 2, 10, and 20 show no discernible pattern in the performance of the QC-based model with respect to the DBN-based architectures. As evident from the remaining fault FDR maps, clear patterns emerge corresponding to the number of hidden units in the RBM layers. The FDRs increase with the number of hidden units. This means that the higher-level abstractions of the input data produced by the DBN-based sub-networks can be better represented by higher dimensions than the original input data. However, this is only true for a few select cases of faults. Based on the FDR heatmaps for each fault, an optimal number of hidden units 26 and 20 in the first and second RBM layer, respectively, are selected for further computational experiments.

Among the 8,000 samples in the faulty dataset, a large portion of the samples are accurately classified as faulty. The average detection rate recorded for the local classifier in the QC-based fault diagnosis model is 99.39%, meaning only 0.61% of the faulty samples remain undetected. A major challenge in developing fault diagnosis models is to adjust the trade-off between the FDRs and the FARs. An increase in FDR is usually accompanied by an increase in the FAR. However, the false positive rate for the proposed QC-based diagnosis model is only 5.25%. With an average FDR of 99.39% and FAR of 5.25%, the performance of the proposed fault diagnosis model is significantly high and can efficiently differentiate faulty process data from normal states of operation.

*Table 2. Fault detection results of the local classifier in the proposed QC-based deep learning model for the TE process case study*

| Fault | FDR (%) | |
|---|---|---|
| | Training | Testing |
| 1 | 100 | 100 |
| 2 | 100 | 100 |
| 3 | 99.1 | 98.0 |
| 4 | 100 | 100 |
| 5 | 100 | 100 |
| 6 | 100 | 100 |
| 7 | 100 | 100 |
| 8 | 100 | 100 |
| 9 | 96.3 | 95.25 |
| 10 | 99.7 | 100 |
| 11 | 100 | 100 |
| 12 | 100 | 100 |
| 13 | 100 | 100 |
| 14 | 100 | 100 |
| 15 | 97.8 | 97.0 |
| 16 | 99.0 | 98.5 |
| 17 | 100 | 100 |
| 18 | 100 | 100 |
| 19 | 99.5 | 99.0 |
| 20 | 100 | 100 |

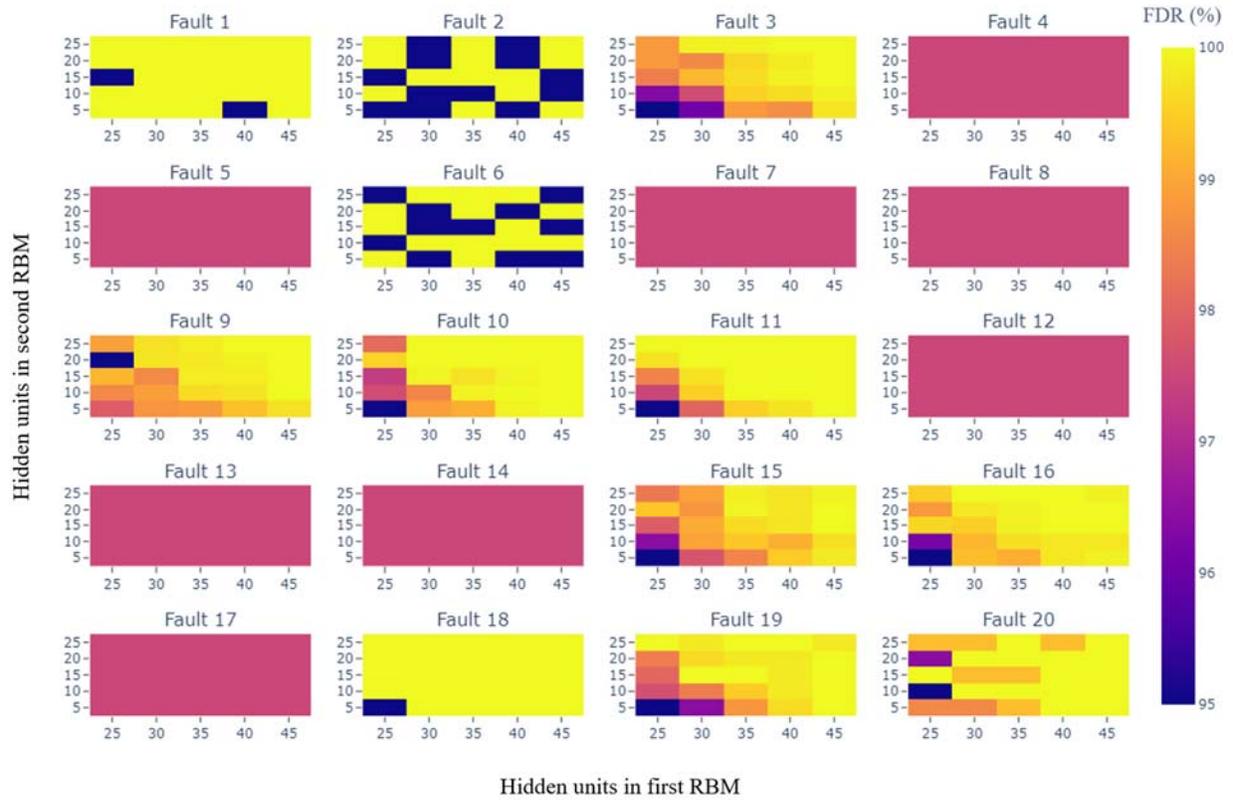

*Figure 12. FDR heatmaps of the DBN sub-networks for the TE process case study*

## 5.3. Fault Identification

Output probabilities that indicate the state of a data sample generated by the repeating sub-network in the QC-based fault diagnosis model are used as individual inputs for the global classification network. For fault identification, it is important to note that the DBN-F sub-network of every repeating sub-network in the QC-based fault diagnosis model is trained with the faulty samples corresponding to its individual fault. The two-dimensional output vector corresponding to individual fault is merged to generate an input vector with 40 dimensions. These input data samples are used to train and validate the performance of the global classifier network. The global classifier network comprises of a fully connected layer with 40 neurons followed by a soft-max layer with 21 neurons. The soft-max layer classifies the input data sample as normal or one of the 20 faulty states. The global classifier is trained with standard back-propagation algorithm with the objective of minimizing the categorical cross-entropy loss.

The diagnosis results for the global classification network that identifies the type of faults produced from the likelihood values obtained with the QC-based fault diagnosis model are reported in Table 3. In the context of fault identification, the FDR metric is modified to compute the classification accuracy for each fault class. To this end, the FDR is computed for each fault type with Eq. (22), where $r$ is the count of a particular faulty state samples that are accurately classified to this state.

$$FDR = \frac{100 \cdot r}{\text{total count of this fault type samples}}\% \quad (22)$$

For comparison purposes, the FDRs of some state-of-the-art data-driven and deep neural network-based approaches are also reported. The diagnosis results obtained for the TE process using PCA [20] and DBN-based fault diagnosis model [34] are also reported in Table 3. As evident from the diagnosis results, PCA does not effectively detect several of the faults in the TE process. The diagnosis rates for faults 3, 9, and 15 are particularly poor. The inability of PCA to take into account the temporal correlations might be contributing to its poor diagnosis performance. This shortcoming is overcome by the DBN-based fault diagnosis model which strongly augments the diagnosis results for a significant portion of the faulty states. The FDRs for faults 3 and 9 improve with the DBN-based model, but fault 15 performs even worse than that of PCA. None of the faults of types 15 and 16 are diagnosed with the DBN-based model in [34]. However, the proposed QC-based fault diagnosis model provides a FDR of 44.9% for fault 15 which is significantly higher than both PCA and DBN-based model proposed in [34]. In the TE process, faults 3, 5, 9, 15, and 16 are particularly hard to detect and usually require significant model tweaking for a mediocre performance improvement. Among the poorly performing fault classes with the QC-based fault diagnosis model, only faults 9, 10, and 11 are random faults. Fault 3 is a step fault introduced in feed temperature in the TE process. The DBN-based model in [34] is unable to detect any fault of types 15 and 16. Most state-of-the-art fault diagnosis models are unable to generalize to multiple faults. A similar trend of generalization to multiple faults is observed with the proposed QC-based fault diagnosis model. Investigating the significance of poorly detected faults on the generalization abilities of the QC-based fault diagnosis model forms the basis for the future scope of this work. On the other hand, the global classifier in the proposed QC-based diagnosis model classifies faults of almost all fault states with a significantly higher accuracy rate than PCA for faults 3 and 9. The resulting FDRs for the rare and hard to detect faults 15 and 16 are higher than those of both PCA and DBN-based models as well. The lowest FDR reported by the QC-based fault diagnosis model is 38.1% for the fault 9. Although, this diagnosis rate for fault 9 is lower than that of the DBN-based model, the diagnostic performance of the proposed model is clearly superior for faults that are rare and hard to detect. Apart from false positives, misclassification of faulty states is also possible and cannot be overlooked as the cost of repairs required due to detection of particular faults could be expensive. The performance of this fault diagnosis model for fault identification can be represented by a confusion matrix which allows the visualization of the accuracy of classification, and misclassification as well. Figure 13 represents the diagnosis results of the global classifier in the QC-based fault diagnosis model in the form of a confusion matrix. The diagonal elements in the matrix are the FDRs for a particular class of samples. The last row in the matrix labeled as normal corresponds to the FARs and have some of the lowest values in the confusion matrix. This confusion matrix can also be used to determine the degree of resemblance between classes of samples. Faults with no similarities whatsoever between other faulty or normal states are relatively easy to diagnose with lower chances of misidentification. Faults 1, 2, 5, 6, and 18 are few such faults with the highest FDR recorded with detection rates as high as 100%.

The low FARs produced by the local classifiers in the QC-based fault diagnosis model are maintained for the diagnosis results of the global classifier network. Several computational experiments are performed for the global classifier network in order to estimate the extent of FARs for each corresponding fault. Figure 14 shows the FARs for each of the 20 faults simulated in the TE process. It should be noted that the highest FAR rate recorded is lower than one percent. Although an average FDR of 82.1% is reported for the DBN-based model [34], it should be noted that this framework is developed specifically for complex chemical processes. Its application to the TE process involves several data preprocessing steps like variable sorting and time length selection that are not considered for this case study performed with the QC-based deep learning model for fault diagnosis. Such preprocessing steps involving feature screening could improve the diagnosis results of the proposed QC-based deep learning model for fault diagnosis. However, to demonstrate the generalization capabilities of the proposed fault diagnosis model, feature screening is not performed for the TE process case study. With an average FDR of 80% and a total average FAR of 1.3%, the proposed QC-based fault diagnosis model can be competitively used against highly targeted state-of-the-art fault diagnosis methods implemented with classical computers and for detection and diagnosis of rare faults in complex process systems.

*Table 3. Comparison between different fault diagnosis models for the TE process case study with respect to fault detection rates for each identified fault*

| Fault | FDR (%) | | |
|:-:|:-:|:-:|:-:|
| | **PCA[20]** | **DBN-based model[34]** | **QC-based model** |
| 1 | 99.88 | 100 | 100 |
| 2 | 98.75 | 99 | 100 |
| 3 | 12.88 | 95 | 51.1 |
| 4 | 100 | 98 | 94.7 |
| 5 | 33.63 | 86 | 100 |
| 6 | 100 | 100 | 100 |
| 7 | 100 | 100 | 95.3 |
| 8 | 98 | 78 | 76.3 |
| 9 | 8.38 | 57 | 38.1 |
| 10 | 60.5 | 98 | 44.6 |
| 11 | 78.88 | 87 | 51.5 |
| 12 | 99.13 | 85 | 81.3 |
| 13 | 95.38 | 88 | 94.7 |
| 14 | 100 | 87 | 86.9 |
| 15 | 14.13 | 0 | 44.9 |
| 16 | 55.25 | 0 | 68.3 |
| 17 | 95.25 | 100 | 92.7 |
| 18 | 90.5 | 98 | 95.6 |
| 19 | 41.13 | 93 | 73.0 |
| 20 | 63.38 | 93 | 89.9 |

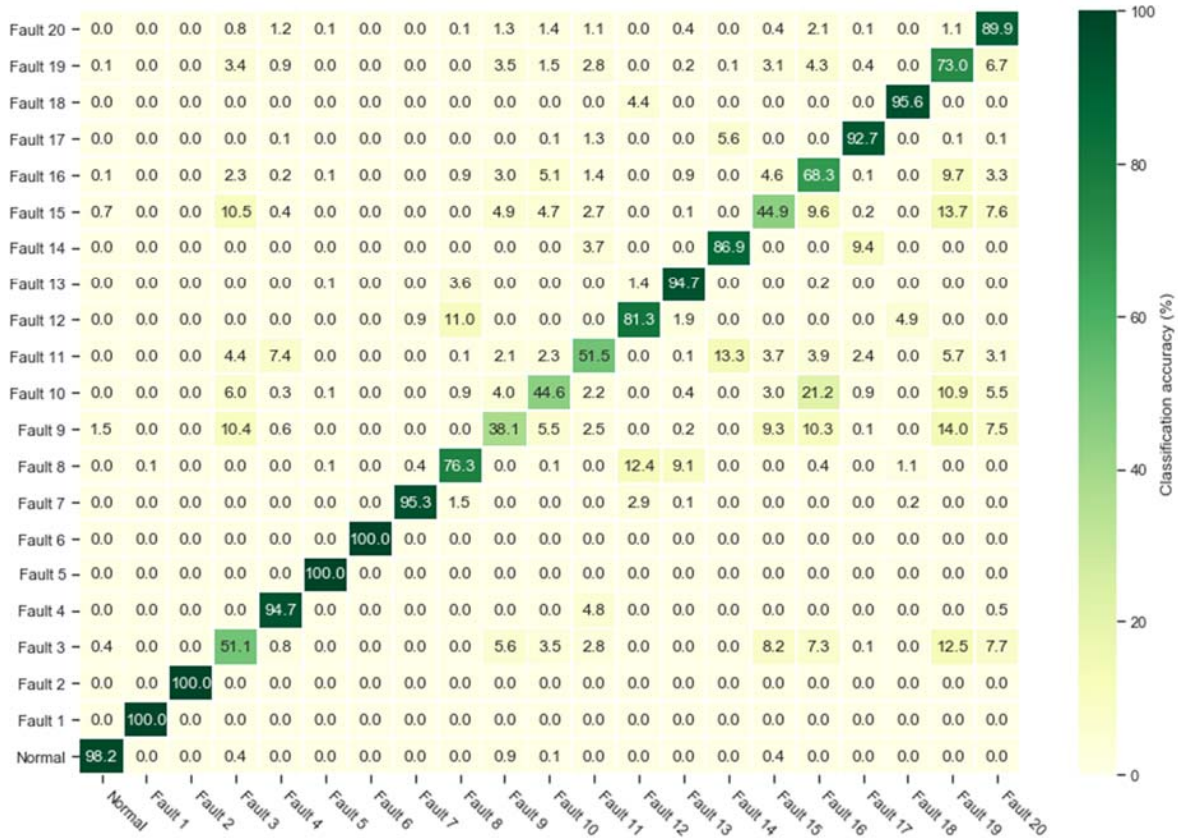

*Figure 13. Confusion matrix for the fault diagnosis results obtained by the global classifier*

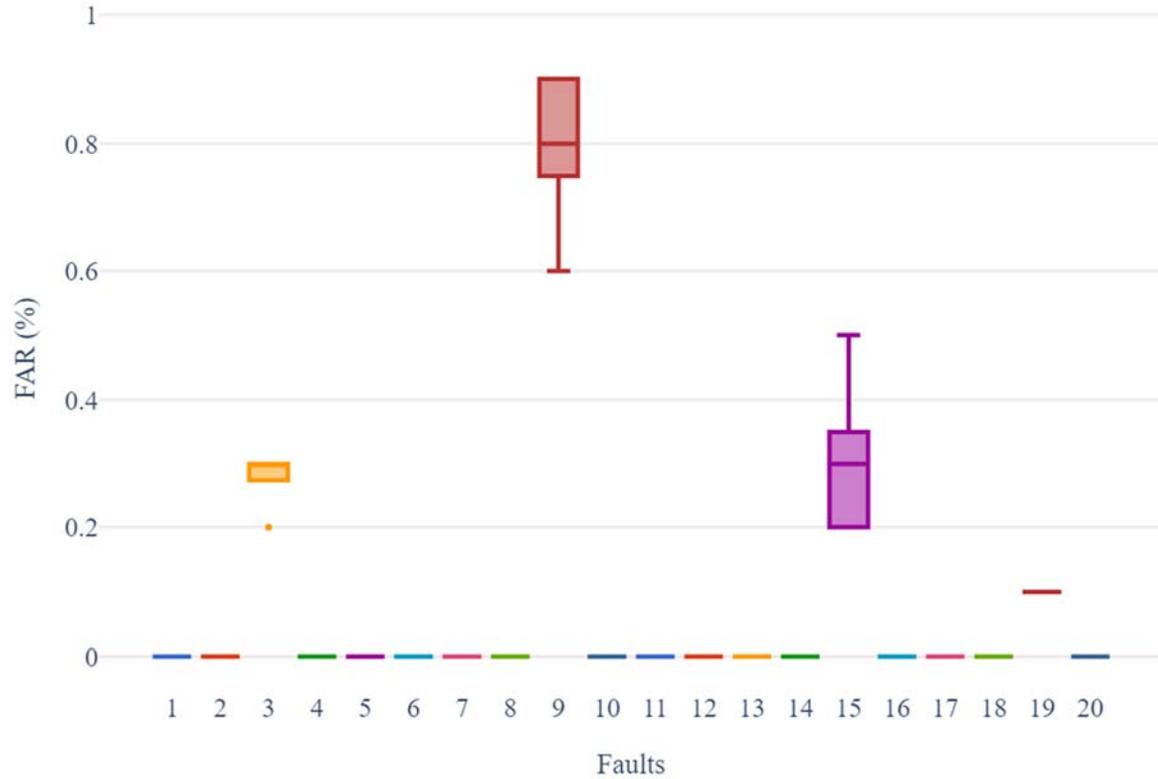

*Figure 14. False alarm rates for the global classifier in the TE process case study*

## 6. Quantum Advantage

Conventional classical learning techniques to train DBNs crudely approximate the log-likelihood gradients of the training data required for the hyper-parameter update rules. The CD-k algorithm more closely approximates the contrastive divergence that is defined as the difference between Kullback-Liebler divergences [54]. It has also been demonstrated that CD-k algorithm does not follow the gradient of any function [64]. Although CD-k converges to the true gradient after infinite reconstruction steps, it is impractical to run the algorithm for an endless time. Other than the approximation limitations, CD-k may take many iterations to converge due to the inherent noise in Gibbs sampling and slow evolution towards the equilibrium probability distribution [17].

Quantum generative training circumvents some challenges put forth by the classical training techniques. For machine learning and deep learning applications, a quantum advantage can be quantified with the computation effort require to achieve a particular model performance. Computation time required could also be considered as a factor in demonstrating the efficiency of quantum inspired techniques over classical techniques. In the case studies, the performance profiles given by the loss curves for the second RBM layer in the DBN-F sub-network can be used to compare performance of classical and quantum training techniques. Loss curves for all faults in the CSTR case study are shown in Figure 15 for both CD-1 algorithm and the quantum sampling-based training approach Similar curves for the TE process case study are also plotted for few faults and are given in Figure 16. These particular representation for faults in the TE process are chosen such that a clear distinction between the classical and quantum techniques can be observed. As seen in the plots, QC-based training algorithm converges faster than the classical CD-1 algorithm. A clear quantum advantage can be perceived with quantum assisted training techniques for the proposed fault diagnosis model. The computation time required to calculate gradients with both quantum and classical techniques is negligible in the case of TE process; therefore, this is not an effective criteria to quantify the superiority of quantum inspired techniques over classical training algorithms. In addition, samples are drawn from an AQC device at each step of the quantum sampling process within 20μs. This sampling time is independent of the size of the RBM network and does not increase with size, unlike classical training techniques. This implies that a computational time advantage could be clearly perceived in case of large networks trained with the quantum generative training process.

The approximation errors for the CD-1 algorithm to train DBNs could have adverse effects on its performance as the size of the RBM sub-networks gets very large. Markov chain based conventional training techniques would not be a feasible choice in such cases either. However, because quantum sampling can draw samples from an underlying approximate Boltzmann distribution that models the joint probability of the RBM, the quantum generative training technique can guarantee an efficient performance. This holds true provided that the size of the RBM energy function does not exceed the scale of current AQC-based computers. As evident from the two case studies of process monitoring in nonlinear complex process systems, the proposed QC-based fault diagnosis model effectively detects faults with significantly higher detection rates and lower false positives. This implies that the proposed fault diagnosis model is a generalized approach and could work for most nonlinear complex process systems with little to no modifications. With the increasing applicability of deep neural networks, a quantum advantage provides an extra edge to such approaches. To this end, it is important to note that computational speed could also contribute towards the quantum advantage as the number of process variables increases. Faster convergence with quantum sampling ensures less computation to achieve the same model performance with that of classical techniques like CD-1 algorithm. High computational speeds coupled with faster convergence can guarantee superior performance of such deep learning models and methods.

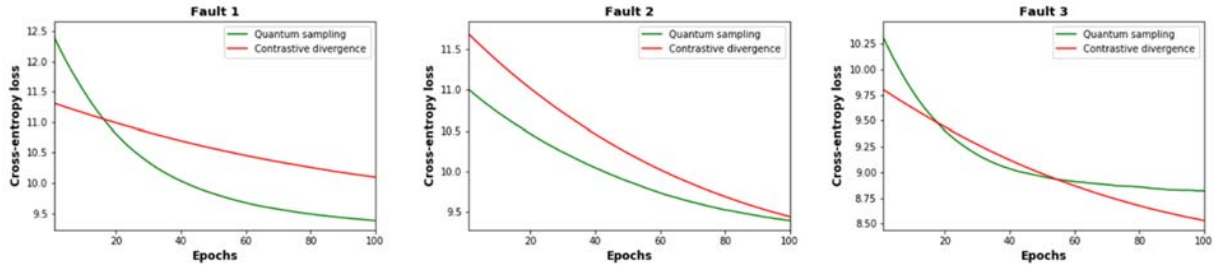

*Figure 15. Loss curves for DBN trained with quantum and classical methods for the CSTR case study*

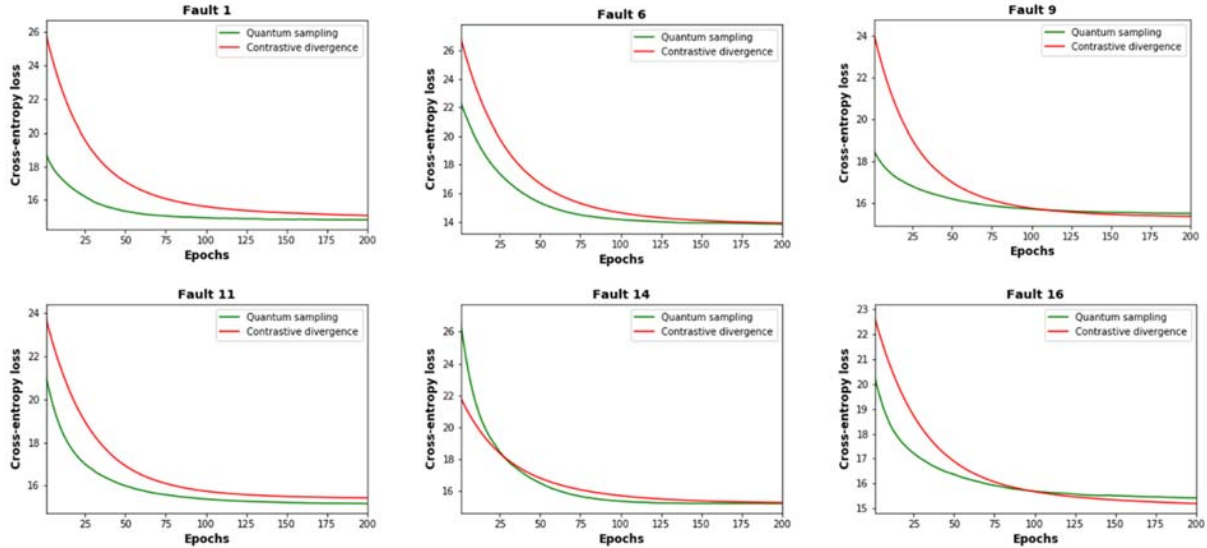

*Figure 16. Loss curves for DBN trained with quantum and classical techniques for the TE process case study*

## 7. Conclusion

In this paper, we proposed a QC-based fault diagnosis model to distinguish faulty states from normal operating states in complex industrial chemical process systems. We integrated quantum assisted generative training with classical discriminative training to detect and diagnose multiple faults introduced in the system. The sampling abilities of AQC computers were exploited to perform quantum generative training for the DBN-based sub-networks present in the proposed QC-based fault diagnosis model. The applicability of this model was demonstrated through two applications on a CSTR and TE process, respectively. The obtained detection and diagnosis results indicated that the proposed QC-based fault diagnosis model clearly outperformed state-of-the-art data-driven approaches and deep neural network based models in most cases. A quantum advantage was also perceived with the quantum generative training while training the DBN-based sub-networks in the proposed fault diagnosis model in contrast to the classical training approaches.